\documentstyle[seceq]{ptptex}

\title{
Generalized Heisenberg's Dynamics
}

\author{
Yoshiharu {\sc Kawamura}\footnote{E-mail:
haru@azusa.shinshu-u.ac.jp} %
}

\inst{
Department of Physics, Shinshu University, Matsumoto 390-8621, Japan
}

\recdate{
}

\abst{
We formulate a dynamical system based on many-index objects.
These objects yield a generalization of the Heisenberg's
equation.
Systems describing harmonic oscillators are given.}

\begin{document}

\maketitle

\section{Introduction}

Recently, a new mechanics has been proposed that is
based on three-index objects\cite{YK},
and its basic structure has been studied from an algebraic point of view\cite{YK2,YK3}.
This mechanics has a counterpart in the canonical structure of classical mechanics
or Nambu mechanics\cite{Nambu},
and can be interpreted as its $\lq$quantum' or $\lq$discretized' version.
It can also be regarded as a generalization of Heisenberg's matrix mechanics,
because a generalization of the Ritz rule 
and that of the Bohr's frequency condition are employed as guiding principles.

The mechanics, 
in which physical variables are $n$-index objects ($n \geq 4$),
was also proposed in Ref. \citen{YK}, but its formulation has not yet been completed.
The purpose of the present paper is the construction of a mechanics 
for mutli-index objects, modeling the dynamical structure of Heisenberg's matrix mechanics.

This paper is organized as follows.
In the next section, we explain the definition of $n$-index objects.
We formulate a dynamical system based on $n$-index objects in $\S$3. 
Section 4 is devoted to conclusions.

\section{Generalized matrices}

\subsection{Definitions}

We state our definition of $n$-index objects
(we refer to them as $n$-th power matrices)\footnote{
Many-index objects have been introduced 
to construct a quantum version of the Nambu bracket.\cite{many-index,Xiong}
The definition of the $n$-fold product we use is the same as that by Xiong.}
and define related terminology.
An $n$-th power matrix is an object with $n$ indices written $A_{l_1 l_2 \cdots l_n}$, 
which is a generalization
of a usual matrix written analogously as $B_{l_1 l_2}$.
We handle $n$-th power matrices that every line has a same size, i.e., 
$N \times N \times \cdots \times N$ matrices,
and treat the elements of an $n$-th power matrix as $c$-numbers 
throughout this paper.

First we define the hermiticity of an $n$-th power matrix by the relation
$A_{l'_1 l'_2 \cdots l'_n} = A_{l_1 l_2 \cdots l_n}^{*}$ for odd permutations among indices
and refer to an $n$-th power matrix possessing hermiticity as a hermitian $n$-th power matrix.
Here, the asterisk indicates complex conjugation.
A hermitian $n$-th power matrix has the relation 
$A_{l'_1 l'_2 \cdots l'_n} = A_{l_1 l_2 \cdots l_n}$ for even permutations among indices.
The components with a same indix, e.g., $A_{l_1 \cdots l_i \cdots l_i \cdots l_n}$,
which is a counterpart of a diagonal part in a hermitian matrix, are real-valued and symmetric
with respect to permutations among indices $\{l_1, \cdots, l_i, \cdots, l_i, \cdots, l_n\}$.
We refer to a special type of hermitian matrix whose components with completely different
indices are vanishing 
as a real normal form or a real normal $n$-th power matrix.
An normal $n$-th power matrix is written 
\begin{eqnarray}
A_{l_1 l_2 \cdots l_n}^{(N)} = \sum_{i < j} \delta_{l_i l_j} 
a_{l_j l_1 \cdots \hat{l}_i \cdots \hat{l}_j \cdots l_n} ,
\label{normal} 
\end{eqnarray}
where the summation is over all pairs among $\{l_1, \cdots, l_n\}$,
the indices with a hat are omitted,
and $a_{l_j l_1 \cdots \hat{l}_i \cdots \hat{l}_j \cdots l_n}$
is symmetric under the exchange of $n-2$ indices 
$\{l_1, \cdots, \hat{l}_i, \cdots, \hat{l}_j, \cdots,l_n\}$.

We define the $n$-fold product of $n$-th power matrices $(A_i)_{l_1 l_2 \cdots l_n}$, 
$(i=1,2,...,n)$ by
\begin{eqnarray}
(A_1 A_2 \cdots A_n)_{l_1 l_2 \cdots l_n} \equiv
\sum_k (A_1)_{l_1 \cdots l_{n-1} k} (A_2)_{l_1 \cdots l_{n-2} k l_n} \cdots (A_n)_{k l_2 \cdots l_n} .
\label{product}
\end{eqnarray}
The resultant $n$-index object, $(A_1 A_2 \cdots A_n)_{l_1 l_2 \cdots l_n}$,
does not necessarily possess hermiticity, even if $(A_i)_{l_1 l_2 \cdots l_n}$s
are hermitian $n$-th power matrices.
Note that the above product is, in general, neither commutative nor associative; for example,
\begin{eqnarray}
&~& (A_1 A_2 \cdots A_n)_{l_1 l_2 \cdots l_n} \neq (A_2 A_1 \cdots A_n)_{l_1 l_2 \cdots l_n} , 
\nonumber \\
&~& (A_1 \cdots A_{n-1} (A_n A_{n+1} \cdots A_{2n-1}))_{l_1 l_2 \cdots l_n} 
\nonumber \\
&~& ~~~~~ \neq ((A_1 \cdots A_{n-1} A_n) A_{n+1} \cdots A_{2n-1})_{l_1 l_2 \cdots l_n} .
\end{eqnarray}
The $n$-fold commutator is defined by
\begin{eqnarray}
&~& [A_1, A_2, \cdots, A_n]_{l_1 l_2 \cdots l_n} 
\nonumber \\
&~& ~~~~~~ \equiv 
\sum_{(i_1, \cdots, i_n)} \sum_k \mbox{sgn}(P) 
(A_{i_1})_{l_1 \cdots l_{n-1} k} (A_{i_2})_{l_1 \cdots l_{n-2} k l_n} 
 \cdots (A_{i_n})_{k l_2 \cdots l_n} ,
\label{commutator}
\end{eqnarray}
where the first summation is over all permutations among the subscripts $\{i_1, \cdots, i_n\}$.
Here, sgn($P$) is $+1$ and $-1$ for even and odd permutations 
among the subscripts $\{i_1, \cdots, i_n\}$, respectively.
If $(A_i)_{l_1 l_2 \cdots l_n}$s are hermitian $n$-th power matrices,
then $i[A_1, A_2, \cdots, A_n]_{l_1 l_2 \cdots l_n}$ is also hermitian . 

\subsection{Properties}

We study some properties on the $n$-fold commutator 
$[A_1, A_2, \cdots, A_n]_{l_1 l_2 \cdots l_n}$.
This commutator is written
\begin{eqnarray}
[A_1, A_2, \cdots, A_n]_{l_1 l_2 \cdots l_n} 
&=& (A_1)_{l_1 l_2 \cdots l_n} \widetilde{(A_2 A_3 \cdots A_n)}_{l_1 l_2 \cdots l_n} 
\nonumber \\
&~& + (-1)^{n-1} (A_2)_{l_1 l_2 \cdots l_n} \widetilde{(A_3 \cdots A_n A_1)}_{l_1 l_2 \cdots l_n}
\nonumber \\
&~& + \cdots 
 + (-1)^{n-1} (A_n)_{l_1 l_2 \cdots l_n} \widetilde{(A_1 A_2 \cdots A_{n-1})}_{l_1 l_2 \cdots l_n}
\nonumber \\
&~& + ([A_1, A_2, \cdots, A_n])^0_{l_1 l_2 \cdots l_n} ,
\label{commutator2}
\end{eqnarray}
where $\widetilde{(A_2 A_3 \cdots A_{n})}_{l_1 l_2 \cdots l_n}$ 
and $([A_1, A_2, \cdots, A_n])^0_{l_1 l_2 \cdots l_n}$ are defined by
\begin{eqnarray}
&~& \widetilde{(A_2 A_3 \cdots A_{n})}_{l_1 l_2 \cdots l_n} 
\nonumber \\
&~& ~~~ \equiv \sum_{(i_2, \cdots, i_n)} \mbox{sgn}(P) \Bigl( (A_{i_2})_{l_1 \cdots l_{n-2} l_n l_n}
(A_{i_3})_{l_1 \cdots l_{n-3} l_n l_{n-1} l_n} \cdots (A_{i_n})_{l_n l_2 \cdots l_{n-1} l_n}
\nonumber \\
&~& ~~~~~~ + (-1)^{n-1} (A_{i_2})_{l_1 \cdots l_{n-3} l_{n-1} l_{n-1} l_n}
(A_{i_3})_{l_1 \cdots l_{n-4} l_{n-1} l_{n-2} l_{n-1} l_n} 
\cdots (A_{i_n})_{l_1 \cdots l_{n-2} l_{n-1} l_{n-1}}
\nonumber \\
&~& ~~~~~~ + \cdots + (-1)^{n-1} (A_{i_2})_{l_1 \cdots l_{n-1} l_1}
(A_{i_3})_{l_1 \cdots l_{n-2} l_{1} l_n} \cdots (A_{i_n})_{l_1 l_1 l_3 \cdots l_n} \Bigr)
\label{widetilde} 
\end{eqnarray}
and 
\begin{eqnarray}
&~& ([A_1, A_2, \cdots, A_n])^0_{l_1 l_2 \cdots l_n} \nonumber \\
&~& ~~~~~~ \equiv 
\sum_{(i_1, \cdots i_n)} \sum_{k \neq l_1, l_2, \cdots, l_n} \mbox{sgn}(P) 
(A_{i_1})_{l_1 \cdots l_{n-1} k} (A_{i_2})_{l_1 \cdots l_{n-2} k l_n} \cdots (A_{i_n})_{k l_2 \cdots l_n} ,
\label{commutator0}
\end{eqnarray}
respectively.

We discuss features of $\widetilde{(A_1 A_2 \cdots A_{n-1})}_{l_1 l_2 \cdots l_n}$. 
It possesses skew-symmetry 
with respect to permutations among indices:
\begin{eqnarray}
&~&  \widetilde{(A_1 A_2 \cdots A_{n-1})}_{l_1 \cdots l_i \cdots l_j \cdots l_n} = 
- \widetilde{(A_1 A_2 \cdots A_{n-1})}_{l_1 \cdots l_j \cdots l_i \cdots l_n} ,
\label{skew-symmetry}
\end{eqnarray}
if $(A_k)_{l_j l_1 \cdots \hat{l}_i \cdots l_n}$s, $(k = 1, ..., n-1)$ 
are symmetric with respect to permutations among $n$-indices 
$\{l_j, l_1, \cdots, \hat{l}_i, \cdots, l_n\}$ as hermitian $n$-th power matrices are.
Let us define an operation for $k$-th anti-symmetric objects $\omega_{l_1 l_2 \cdots l_k}$ by
\begin{eqnarray}
(\delta{\omega})_{l_0 l_1 l_2 \cdots l_k} \equiv 
\sum_{i=0}^{k} (-1)^{i} {\omega}_{l_0 l_1 \cdots \hat{l}_i \cdots l_k}  .
\label{delta}
\end{eqnarray}
Here the operator $\delta$ is regarded as a coboundary operator 
that changes $k$-th antisymmetric objects into $(k+1)$-th objects,
and this operation is nilpotent, i.e. $\delta^2(*) =0$.\footnote{
See Ref.~\citen{cohom} for textbooks with respect to cohomology.}
If the $\omega_{l_1 l_2 \cdots l_k}$ satisfies
the cocycle condition: $(\delta \omega)_{l_0 l_1 l_2 \cdots l_k} = 0$,
it is called a $k$-cocycle.
We give an example of a solution for the cocycle condition: 
$(\delta \widetilde{(A_1 A_2 \cdots A_{n-1})})_{l_0 l_1 l_2 \cdots l_n} = 0$. 
When one of $A_l$s is an arbitrary hermitian $n$-th power matrix and all the rest
have components in the form $(A_l)_{l_j l_1 \cdots \hat{l}_i \cdots l_n}
 = \sum_{l_k \neq \hat{l}_i} (a_l)_{l_k}$,
the $\widetilde{(A_1 A_2 \cdots A_{n-1})}_{l_1 l_2 \cdots l_n}$ is written
\begin{eqnarray}
\widetilde{(A_1 A_2 \cdots A_{n-1})}_{l_1 l_2 \cdots l_n} 
= \sum_{i=1}^{n} (-1)^{i-1} \gamma_{l_1 \cdots \hat{l}_i \cdots l_n}
\equiv (\delta \gamma)_{l_1 l_2 \cdots l_n} ,
\label{coboundary}
\end{eqnarray}
where $\gamma_{l_1 l_2 \cdots l_{n-1}}$ is an $(n-1)$-th antisymmetric objects.
Then,
the $n$-th antisymmetric object
$\widetilde{(A_1 A_2 \cdots A_{n-1})}_{l_1 l_2 \cdots l_n}$
automatically satisfies the cocycle condition:
\begin{eqnarray}
(\delta{\widetilde{(A_1 A_2 \cdots A_{n-1})}})_{l_0 l_1 l_2 \cdots l_n} =
(\delta^2 \gamma)_{l_0 l_1 \cdots \hat{l}_i \cdots l_n} = 0 
\label{coboundary2}
\end{eqnarray}
due to the nilpotency for coboundary operations.
This type of solution is called an $n$-coboundary. 

We can show the following relations on the $n$-fold commutator
from the above expressions and properties.
\begin{enumerate} 
\item For arbitrary $n$-th power hermitian matrices $A_j$, 
$[A_1, \cdots, A_{n-1}, \Delta]_{l_1 l_2 \cdots l_n} = 0$
with $\Delta_{l_1 l_2 \cdots l_n} = \prod_{i < j} \delta_{l_i l_j}$.
Here the product is over all pairs among indices.

\item For arbitrary normal $n$-th power matrices $A_j^{(N)}$,
the $n$-fold commutator among $A$ and $A_j^{(N)}$ is
given by 
\begin{eqnarray}
[A, A_1^{(N)}, \cdots, A_{n-1}^{(N)}]_{l_1 l_2 \cdots l_n} 
= A_{l_1 l_2 \cdots l_n} \widetilde{(A_1^{(N)} \cdots A_{n-1}^{(N)})}_{l_1 l_2 \cdots l_n}.
\label{formula1}
\end{eqnarray}

\item The $n$-fold commutator among
arbitrary normal $n$-th power matrices $A_i^{(N)}$
is vanishing; 
\begin{eqnarray}
[A_1^{(N)}, A_2^{(N)}, \cdots, A_{n}^{(N)}]_{l_1 l_2 \cdots l_n} = 0.
\label{formula2}
\end{eqnarray}

\item If $\widetilde{(B_1^{(N)} B_2^{(N)} \cdots B_{n-1}^{(N)})}_{l_1 l_2 \cdots l_n}$ is an $n$-cocycle
for normal $n$-th power matrices $B_j^{(N)}$,
the fundamental identity holds:
\begin{eqnarray}
&~& [[A_1, \cdots, A_n], B_1^{(N)}, \cdots, B_{n-1}^{(N)}]_{l_1 l_2 \cdots l_n} 
\nonumber\\
&~& ~~~ =  \sum_{i=1}^{n} [A_1, \cdots, [A_i, B_1^{(N)}, 
\cdots, B_{n-1}^{(N)}], \cdots, A_n]_{l_1 l_2 \cdots l_n} .
\label{fundamental}
\end{eqnarray}
\end{enumerate}

\section{Dynamical system}

In this section, we employ a generalization of the Ritz rule 
and that of the Bohr's frequency condition as guiding principles, and construct
a generalization of Heisenberg's matrix mechanics based on hermitian $n$-th power matrices.

\subsection{Framework}

The time-dependent variables are hermitian $n$-th power matrices given by
\begin{eqnarray}
(V_{\alpha}(t))_{l_1 l_2 \cdots l_n} 
=  (V_{\alpha})_{l_1 l_2 \cdots l_n} e^{i\Omega_{l_1 l_2 \cdots l_n}t} ,
\label{variable1}
\end{eqnarray}
where the angular frequency $\Omega_{l_1 l_2 \cdots l_n}$ has the properties
\begin{eqnarray}
&~& \Omega_{l'_1 l'_2 \cdots l'_n} = \mbox{sgn}(P) \Omega_{l_1 l_2 \cdots l_n} , 
\label{antisymmetry}\\
&~& (\delta \Omega)_{l_0 l_1 l_2 \cdots l_n} \equiv 
\sum_{i=0}^{n} (-1)^{i} \Omega_{l_0 l_1 \cdots \hat{l}_i \cdots l_n} = 0 .
\label{cocycle2}
\end{eqnarray} The $\Omega_{l_1 l_2 \cdots l_n}$ is 
regarded as an $n$-cocycle, from the equation (\ref{cocycle2}).
This equation is a generalization of the Ritz rule,\footnote{
The Ritz rule is given by $\Omega_{l_1 l_3} = \Omega_{l_1 l_2} + \Omega_{l_2 l_3}$ in quantum mechanics, 
where $\Omega_{l_i l_j}$ is the angular frequency of radiation from an atom.}
and it is required from a consistency for the time evolution of a system as will be shown.
Notice that the $n$-fold product,
$(V_{\alpha_1} V_{\alpha_2} \cdots V_{\alpha_n})_{l_1 l_2 \cdots l_n} e^{i\Omega_{l_1 l_2 \cdots l_n}t}$,
takes the same form as (\ref{variable1}), with the relation (\ref{cocycle2}). 

The time-independent variables are real normal $n$-th power matrices given by
\begin{eqnarray}
(U_a)_{l_1 l_2 \cdots l_n} =  \sum_{i < j} \delta_{l_i l_j} 
(u_a)_{l_j l_1 \cdots \hat{l}_i \cdots \hat{l}_j \cdots l_n} .
\label{variable2}
\end{eqnarray}
These variables are conserved quantities, and
are regarded as generators of a symmetry transformation.

Next we discuss the time evolution of physical variables.
It is given as a symmetry transformation 
if the physical system is closed.
In our mechanics, it is expected 
that real normal $n$-th power matrices generate such a transformation.
We refer to them as $\lq$Hamiltonians' $H_A$ $(A=1, ... ,n-1)$ .
Hamiltonians are functions of physical variables: $H_A = H_A(V_{\alpha}(t), U_a)$.
By analogy with Heisenberg's matrix mechanics, 
we require that the $(V_{\alpha}(t))_{l_1 l_2 \cdots l_n}$s should yield the generalization of 
the Heisenberg equation:
\begin{eqnarray}
 {d \over dt}(V_{\alpha}(t))_{l_1 l_2 \cdots l_n}  
 = {1 \over i\hbar^{(n)}} [V_{\alpha}(t), H_1, \cdots, H_{n-1}]_{l_1 l_2 \cdots l_n} ,
\label{GH-eq}
\end{eqnarray}
where $\hbar^{(n)}$ is a new physical constant.
The left-hand side of (\ref{GH-eq}) is written
\begin{eqnarray}
{d \over dt}(V_{\alpha}(t))_{l_1 l_2 \cdots l_n}  
= i \Omega_{l_1 l_2 \cdots l_n} (V_{\alpha}(t))_{l_1 l_2 \cdots l_n} ,
\label{LHS}
\end{eqnarray}
by definition (\ref{variable1}).
On the other hand, the right-hand side is written 
\begin{eqnarray} 
&~& {1 \over i\hbar^{(n)}} [V_{\alpha}(t), H_1, \cdots, H_{n-1}]_{l_1 l_2 \cdots l_n} 
\nonumber\\
&~& ~~~~~~ = {1 \over i\hbar^{(n)}} (\widetilde{H_1 \cdots H_{n-1}})_{l_1 l_2 \cdots l_n}
 (V_{\alpha}(t))_{l_1 l_2 \cdots l_n} ,
\label{RHS}
\end{eqnarray}
by use of the formula (\ref{formula1}).
{}From equations (\ref{LHS}) and (\ref{RHS}), we obtain the relation
\begin{eqnarray} 
\hbar^{(n)} \Omega_{l_1 l_2 \cdots l_n} 
= - (\widetilde{H_1 \cdots H_{n-1}})_{l_1 l_2 \cdots l_n} .
\label{GB-rel}
\end{eqnarray}
This relation is a generalization of Bohr's frequency condition.\footnote{
The Bohr's frequency condition is given by 
$\hbar \Omega_{l_1 l_2} = - \widetilde{H}_{l_1 l_2} = E_{l_1} - E_{l_2}$ in quantum mechanics, 
where $E_{l}$ is the energy eigenvalue of an atom.} 

Let us make a consistency check for the time evolution of a system.
By definition, an arbitrary normal $n$-th power matrix $A^{(N)}$ (and the time-independent part of
$V_{\alpha}(t)$)
should be a constant of motion, and it is verified by use of the equation of motion:
\begin{eqnarray} 
\frac{d}{dt} (A^{(N)})_{l_1 l_2 \cdots l_n} 
= \frac{1}{i \hbar^{(n)}} [A^{(N)}, H_1, \cdots, H_{n-1}]_{l_1 l_2 \cdots l_n} = 0 ,
\end{eqnarray}
where the formula (\ref{formula2}) is used. 
Since the Hamiltonians are real normal $n$-th power matrices, they are conserved quantities.
The $n$-fold commutator, $i[V_1(t), \cdots , V_n(t)]$ should satisfy the fundamental
identity including the Hamiltonians:
\begin{eqnarray}
&~& [i[V_1(t), \cdots, V_n(t)], H_1, \cdots, H_{n-1}]_{l_1 l_2 \cdots l_n} 
\nonumber\\
&~& ~~~ =  \sum_{i=1}^{n} i[V_1(t), \cdots, [V_i(t), H_1, 
\cdots, H_{n-1}], \cdots, V_n(t)]_{l_1 l_2 \cdots l_n} 
\label{fundamental2}
\end{eqnarray}
from the requirement that the derivation rule for the time should hold such that
\begin{eqnarray}
\frac{d}{dt}i[V_1(t), \cdots, V_n(t)]_{l_1 l_2 \cdots l_n} 
 =  \sum_{i=1}^{n} i[V_1(t), \cdots, \frac{d}{dt}V_i(t), \cdots, V_n(t)]_{l_1 l_2 \cdots l_n} .
\label{derivation}
\end{eqnarray}
The fundamental identity (\ref{fundamental2}) holds 
in the case that the $\Omega_{l_1 l_2 \cdots l_n}$ is an $n$-cocycle from 
the 4-th relation on the $n$-fold commutator.

\subsection{Examples}

We study the simple example of a harmonic oscillator whose variables are two kinds of hermitian 
$n$-th power matrices given by
$\xi(t)_{l_1 l_2 \cdots l_n} = \xi_{l_1 l_2 \cdots l_n} e^{i\Omega_{l_1 l_2 \cdots l_n}t}$ and
$\eta(t)_{l_1 l_2 \cdots l_n} = \eta_{l_1 l_2 \cdots l_n} e^{i\Omega_{l_1 l_2 \cdots l_n}t}$. 
Here, each of the indices $l_i$ runs from 1 to $n$.
The coefficients $\xi_{l_1 l_2 \cdots l_n}$ and $\eta_{l_1 l_2 \cdots l_n}$ are given by
\begin{eqnarray}
\xi_{l_1 l_2 \cdots l_n} =  \sqrt{{\hbar^{(n)} \over 2m\Omega}} |\varepsilon_{l_1 l_2 \cdots l_n}| , ~~
\eta_{l_1 l_2 \cdots l_n} = {1 \over i}\sqrt{{m \Omega \hbar^{(n)} \over 2}} \varepsilon_{l_1 l_2 \cdots l_n} ,
\label{xieta}
\end{eqnarray}
where the quantity $m$ in the square root represents a mass, $\Omega = |\Omega_{1 2 \cdots n}|$
and $\varepsilon_{l_1 l_2 \cdots l_n}$ is the $n$-dimensional Levi-Civita symbol.

If $\Omega_{l_1 l_2 \cdots l_n} = - \Omega \varepsilon_{l_1 l_2 \cdots l_n}$,
we obtain the equations of motion describing the harmonic oscillator:
\begin{eqnarray}
&~& {d \over dt} \xi(t)_{l_1 l_2 \cdots l_n} = {1 \over m} \eta(t)_{l_1 l_2 \cdots l_n} ,
\label{xi-eq} \\
&~& {d \over dt} \eta(t)_{l_1 l_2 \cdots l_n} = - m \Omega^2 \xi(t)_{l_1 l_2 \cdots l_n} . 
\label{eta-eq}
\end{eqnarray}
The hamiltonians $H_A$ satisfy the following relation
\begin{eqnarray} 
\hbar^{(n)} \Omega \varepsilon_{l_1 l_2 \cdots l_n} 
= (\widetilde{H_1 \cdots H_{n-1}})_{l_1 l_2 \cdots l_n} 
\label{GB-rel2}
\end{eqnarray}
from the requirement that physical variables should yield the generalized Heisenberg's
equation (\ref{GH-eq}).
As an example of $H_A$s, we have
\begin{eqnarray}
 (H_1)_{n~n~1 \cdots n-2} = \hbar^{(n)} \Omega , ~~
 (H_B)_{n~n~1 \cdots \hat{i} \cdots n-1} = \delta_{n-i~B} ,
\label{HAs} 
\end{eqnarray}
where $H_A$s are symmetric with respect to permutations among indices, 
other components of $H_A$s are vanishing, and $B$ runs from 2 to $n-1$.
There is the algebraic relation among $\xi(t)$, $\eta(t)$ and $H_A$s:
\begin{eqnarray} 
(H_1)_{l_1 l_2 \cdots l_n} = i \Omega [\xi(t), \eta(t), H_2, \cdots, H_{n-1}]_{l_1 l_2 \cdots l_n} .
\label{algebra}
\end{eqnarray}

Finally we give an example of $\lq$$n$-plet', $(V_{\alpha}(t))_{l_1 l_2 \cdots l_n}$
$(\alpha = 1, \cdots, n)$, where each of the indices $l_i$ runs from 1 to $n^2$.
The components of $V_{\alpha}(t)$s are defined by
\begin{eqnarray} 
(V_{\alpha}(t))_{l_1 l_2 \cdots l_n} \equiv 
\left\{
\begin{array}{ll}
\eta(t)_{l_1 l_2 \cdots l_n} & \quad\mbox{for}~~ 
 l_i = (\alpha -1)n + 1, (\alpha -1)n + 2, \cdots, \alpha n\\
\xi(t)_{l_1 l_2 \cdots l_n} & 
\quad\mbox{for}~~ l_i = \alpha n + 1, \alpha n + 2, \\
~~ & ~~~~~~~~~~~~~~ \cdots, (\alpha + 1) n ~~ (\mbox{mod}~~ n^2)\\
(\zeta_{n-j})_{l_1 l_2 \cdots l_n} & 
\quad\mbox{for}~~ l_i = (\alpha + j -1)n + 1, (\alpha + j -1)n + 2, \\
~~ & ~~~~~~~~~~~~~~ \cdots, (\alpha + j) n ~~ (\mbox{mod}~~ n^2) ,
\end{array}\right.
\label{Valpha}
\end{eqnarray}
where $(\zeta_{n-j})_{l_1 l_2 \cdots l_n}$ is the $n \times n \times \cdots \times n$ matrices 
whose non-vanishing components are given by
\begin{eqnarray} 
(\zeta_{n-j})_{kn~kn~(k-1)n + 1 \cdots \widehat{(k-1)n + i} \cdots kn-1} = \delta_{n-i~n-j+1} .
\label{zeta}
\end{eqnarray}
Here, $i$ and $j$ run from 2 to $n-1$ and $\zeta_{n-j}$s are 
symmetric with respect to permutations among indices
$\{kn~kn~(k-1)n + 1 \cdots \widehat{(k-1)n + i} \cdots kn-1\}$ for $k = 1, \cdots, n$.
We find that the time-dependent components in $V_{\alpha}(t)$s
yield the equation of motion of harmonic oscillators 
under the Hamiltonians whose non-vanishing components 
are given by
\begin{eqnarray}
(H_1)_{kn~kn~(k-1)n + 1 \cdots kn-2} =  (-1)^{k(n-1)} \hbar^{(n)} \Omega ,
\label{H1}\\
(H_B)_{kn~kn~(k-1)n + 1 \cdots \widehat{(k-1)n + i} \cdots kn-1} = \delta_{n-i~B} ,
\label{HB}
\end{eqnarray}
where $B$ runs from 2 to $n-1$ and $k$ runs from 1 to $n$.
The $H_A$s are symmetric with respect to permutations among indices.
The $(V_{\alpha}(t))_{l_1 l_2 \cdots l_n}$s satisfy the following algebra:
\begin{eqnarray} 
[V_1(t), V_2(t), \cdots, V_n(t)]_{l_1 l_2 \cdots l_n} = - i \hbar^{(n)} (J^{(N)})_{l_1 l_2 \cdots l_n} ,
\label{algebra2}
\end{eqnarray}
where $(J^{(N)})_{l_1 l_2 \cdots l_n}$ is the real normal
$n^2 \times n^2 \times \cdots \times n^2$ matrices
whose non-vanishing components are given by
\begin{eqnarray}
(J^{(N)})_{kn~kn~(k-1)n + 1 \cdots kn-2} =  (-1)^{k(n-1)} .
\label{JN}
\end{eqnarray}
In both cases,
the $n+1$ variables, $(\xi(t), \eta(t), H_A)$ or
$(V_{\alpha}(t), J^{(N)})$,
form a closed algebra for the $n$-fold commutator,
which is regarded as a generalization of spin algebra\cite{Filippov,Hoppe,Xiong,YK4}.

\section{Conclusions}

We have given our definition of $n$-index objects ($n$-th power matrices)
and their algebraic properties,
and formulated a dynamical system based on hermitian $n$-th power matrices.
The basic structure of our mechanics is summarized as follows. 
For hermitian $n$-th power matrices $(V_{\alpha}(t))_{l_1 l_2 \cdots l_n}
= (V_{\alpha})_{l_1 l_2 \cdots l_n} e^{i\Omega_{l_1 l_2 \cdots l_n}t}$,
their time evolution is regarded as the symmetry transformation generated by
the Hamiltonians $(H_A)_{l_1 l_2 \cdots l_n}$, such that 
$i \hbar^{(n)} \delta (V_{\alpha}(t))_{l_1 l_2 \cdots l_n} 
= [V_{\alpha}(t), H_1, \cdots, H_{n-1}]_{l_1 l_2 \cdots l_n} \delta t$,
which is the generalization of the Heisenberg's equation.
The Hamiltonians $(H_A)_{l_1 l_2 \cdots l_n}$ are real normal forms 
where $(\widetilde{H_1 \cdots H_{n-1}})_{l_1 l_2 \cdots l_n}$ 
satisfies $n$-cocycle condition.
There is  the relation among $\Omega_{l_1 l_2 \cdots l_n}$ and $H_A$s such that
$\hbar^{(n)} \Omega_{l_1 l_2 \cdots l_n} = - (\widetilde{H_1 \cdots H_{n-1}})_{l_1 l_2 \cdots l_n}$.
An arbitrary normal $n$-th power matrix, $A^{(N)}_{l_1 l_2 \cdots l_n}$, is a constant of motion; i.e., 
$i \hbar^{(n)} dA^{(N)}_{l_1 l_2 \cdots l_n}/dt 
= [A^{(N)}, H_1, \cdots, H_{n-1}]_{l_1 l_2 \cdots l_n} = 0$.
There are simple systems of harmonic oscillators 
described by hermitian $n$-th power matrices.

Our mechanics is regarded as the generalization of 
Heisenberg's matrix mechanics because it reduces to Heisenberg's matrix mechanics
in case with $n=2$.
In quantum mechanics, a matrix element $A_{l_1 l_2}$ is interpreted 
as a probability amplitude between the state 
labeled by $l_1$ and that labeled by $l_2$.
A similar interpretation for an $n$-th power matrix element ($n \geq 3$), however,
is not yet known, and
it is not clear whether many-index objects is applicable to real physical systems.\footnote{
In Ref. \citen{YK4}, a generalization of spin algebra using three-index objects has been proposed
and the connection between a triple commutation relation and an uncertainty relation
has been discussed.}
It would be worth while to explore physical meaning of many-index objects
based on generalized spin variables.

\section*{Acknowledgements}
This work was supported in part by Scientific Grants from the Ministry of Education
and Science, Grant No. 13135217 and Grant No. 15340078.

\end{document}